\magnification=\magstep1
\parskip=4pt
\font\huge=cmbx10 scaled \magstep2

\font\sc=cmcsc10

\def\heading#1{\goodbreak\bigskip\centerline{\sc #1}\smallskip}
\def\subheading#1{\medskip\centerline{\it #1}\smallskip}
\def\etal{{\it et al.\ }}
\def\\{\hfil\break}

\def\q#1{\hbox to 1.25truein{\hfil #1 \hfil}}
\newcount\qno
\def\Eqn{(\the\qno)}
\def\nxt{\global\advance\qno by1 \eqno\Eqn}
\def\wt{$w(\theta)$}
\vbox to 1.5truein{}

\centerline{\huge Strong Angular Clustering of Very Blue
Galaxies:}     \smallskip
\centerline{\huge Evidence of a Low Redshift  Population}
          \bigskip
\centerline{\bf
     Stephen D. Landy$^1$, Alexander S. Szalay$^{2,3}$
     and David C. Koo$^4$}

\medskip\bigskip\bigskip

\noindent {\rm
\item{1.} Carnegie Observatories, 813 Santa Barbara St.,Pasadena,
CA 91101 \item{2.} Dept. of Physics and Astronomy, The Johns
Hopkins University,\\    34th \& Charles St., Baltimore, MD 21218
\item{3.} Dept. of Physics, E\"otv\"os University, Budapest,
H-1088 \item{4.} UCO/Lick Observatory and Board of Studies in
Astronomy and       Astrophysics,\\
     University of California, Santa Cruz, CA 95064

}
\bigskip

\centerline{\huge\bf Abstract}

     We have studied galaxy two-point angular correlations as a
function of color using 4-m plate photometry in two independent
fields. Each field consists of over 2900 galaxies with magnitudes
$20<B_J <23.5$ in an area of approximately 750 arcmin$^2$ . We
find that the autocorrelation amplitude of the bluest 15\% of
galaxies is surprisingly strong, with a relative increase in
clustering amplitude of a factor of 6 over that of the complete
data set, while exhibiting a power law slope consistent with the
canonical value of $-0.8$. These very blue galaxies are also
found to be weakly correlated with galaxies of median color and
marginally anti-correlated with the reddest subset. These
correlation properties are incompatible with existing simple models
of the galaxy distribution; they suggest that a significant
fraction, more than 50\%, of these very blue galaxies are a faint population
which lie at nearby redshifts $z<0.3$.

\bigskip

Subject headings:

Cosmology: Large Scale Structure

Galaxies: Clustering, Distances and Redshifts, Evolution,
Luminosity  Function

\bigskip

\heading{ 1. Introduction}

The nature of faint field galaxies and their high surface density
at faint magnitudes remain a mystery despite extensive new
observations over the last 15 years (see Koo and Kron 1992, Lilly
1993, and Koo 1996 for recent reviews). Besides fainter number
counts and colors, and more recently the morphology of faint
galaxies from Hubble Space Telescope observations (Griffiths {\it
et al.} 1994), the key new observations have been redshifts to
limits as faint as $B_J=24$ for 12 galaxies (Cowie, Songaila, \&
Hu 1991), and 73 galaxies (Glazebrook \etal 1995) and to $I=22$ for
almost 600 galaxies (Lilly \etal 1995).  The detailed conclusions
of these deep redshift surveys remain at odds, though all agree
that substantial evolution has occurred in the luminosity
function of field galaxies (Lilly \etal 1995, Colless 1995).
Whether such changes are caused by a new population, substantial
mergers, or enhanced star-formation among dwarfs has yet to be
resolved. Moreover, both of the larger surveys remain incomplete
at the 20\% to 27\% level, with the bulk of the unidentified
galaxies being bluer than average. This level of incompleteness is
high enough to make a significant difference in understanding the
nature of faint galaxies, especially if the bulk of the
unidentified population is either at high or low redshifts.

    This paper places additional constraints on the nature of the
bluest galaxies by using another observable property of faint
galaxies, namely, their angular clustering behavior. Although the
approach of using angular correlation functions to study galaxy
evolution is not new (Koo \& Szalay 1984) and has been applied to
several samples of faint galaxies, at different wavelengths, or
for gross cuts in color (Efstathiou \etal 1991, Bernstein \etal
1994, Infante \& Pritchet 1995, Neuschaefer \& Windhorst 1995),
none of the analyses so far have explored the variation of
clustering among galaxies of extreme colors. Here we
use the angular correlation function as a tool to measure, or at
least constrain, the redshift distribution of such galaxies.

     The two-point angular correlation function $w(\theta)$ is
well  known and defined by the joint probability $\delta P$ of
finding galaxies  in both of the elements of solid angle
$\delta\Omega_1$ and $\delta\Omega_2$ placed at an angular
separation $\theta$,
$$\delta P = N^2[1+w(\theta)]\delta\Omega_1\delta\Omega_2 \nxt$$
where $N$ is the mean surface density of galaxies. The
observations are  usually close to a power law,
     $$w(\theta)=A_w \theta^{1-\gamma},\nxt$$
with amplitude $A_w$ and $\gamma \sim 1.8$ (see Peebles 1980). As
an  estimator for this function we will use that described in
Landy \&  Szalay (1993) for its improved error properties.

     The scaling of angular correlations as a function of
limiting magnitude is well understood: as the limiting magnitude
of a survey increases, there are more projected pairs that weakens
the correlations (Peebles 1980), and the amplitude is proportional to
an integral over the second moment of $dn/dz$, the redshift
distribution of the galaxies,
$$ A_w \propto \int dz\ c(z) \left({dn\over dz}\right)^2\Biggl/
     \left[\int dz\  \left({dn\over dz}\right)\right]^2, \nxt
$$
where $c(z)$ depends on evolution, curvature, and $\gamma$ (see
e.g.  Koo \& Szalay 1984).
Less is known about changes due to color selection.
Differences in the measured clustering as a function of angle
\wt, for subsamples such as those chosen by color, can occur for
two reasons: each subsample may be dominated by a different
population with different clustering properties, and/or different
redshift distributions. This degeneracy between intrinsic
clustering and redshift distributions can be broken by
considering other information such as the overall shape of the
correlation function, cross-correlations with known populations,
and number counts. Correlation functions are normally not thought
of as a probe of redshift distributions, but as pointed out
earlier (Koo and Szalay 1984), angular correlation functions
provide powerful constraints on this distribution (and hence
evolution and cosmology), which is particularly valuable in the
absence of spectroscopic redshifts for very faint samples, for
example, to $I > 26$.  Thus the method serves as an independent
check or constraint on models that attempt to describe the more
easily measured colors and counts of faint galaxies.

     Since early-type (E/SO/Sa) galaxies have stronger than
average intrinsic clustering (Giovanelli, Haynes, \& Chincarini
1986), the {\it a priori} expectation is for red galaxies to have
correlation amplitudes higher than that of an average field
sample.  In contrast, late-type, blue galaxies today appear to be
more weakly clustered than an average field sample (Nicotra,
Abadi, \& Lambas 1993). Some groups have proposed that such
galaxies were brighter in the past because of enhanced star formation
(Broadhurst, Ellis, \& Shanks 1988). If so, their redshift
distributions will be more extended than without this luminosity
increase. In this case, one might expect a further reduction in
clustering amplitude for blue galaxies observed to fainter
limiting magnitudes. It has also been recently reported that low
surface brightness galaxies are more weakly clustered than the
general population (Neuschaefer \& Windhorst 1995), also expected
in a CDM dominated universe (Efstathiou 1995).

     To compare these expectations with observations, we have
calculated the correlation functions of subsets of galaxies
chosen as a function of color: $U-B_J$, $U-R_F$, and $B_J-R_F$.
We start with extreme subsets of either red or blue objects and
then increase the range to include objects of less extreme color.
Also calculated is the correlation function for samples selected
in bands of color as well as the cross correlation between
samples selected in different bands. We report results on
$U-R_F$, as this color gives the largest wavelength range to
discriminate color-dependent effects.  In addition, this is the
combination for which the galaxy colors are least explicitly dependent on their
redshift, and thus most correlated with intrinsic color
(Koo 1986). The correlation amplitude dependence on $U-B_J$ and
$B_J-R_F$ is similar to $U-R_F$, although the enhancements were
weaker.

\heading{2. Data}

     The photometry data are from two fields of approximately 750
arcmin$^2$ each and based on photographic plates taken at the
prime focus of the Mayall 4m telescope at Kitt Peak National
Observatory. One field SA57 (1306+2939) is at the North Galactic
Pole; the other field SA68 (0015+1537) has galactic coordinates
of $l=-111$ and $b=-46$ and is near Perseus-Pisces. Details of
the observations and data reduction techniques have been
published elsewhere (Kron 1980, Koo 1986). Here it is sufficient
to note that the plates were digitized using the Berkeley PDS
microdensitometer; that areas in the fields around bright stars
as well as bad pixel lines in the data reduction have been
excised from the fields; that both fields are considered complete
to $B_J$ = 23.5, and finally that two plates were used in each
bandpass with only multiple detections included in the catalogs
to decrease the incidence of artifacts. We have selected galaxies
in the magnitude range $20<B_J <23.5$, with 2922 galaxies in SA57
and 2999 galaxies in SA68.

     The correlation properties of the same data have previously
been analyzed, but only as a function of $B_J$ magnitude (Koo \&
Szalay 1984). This earlier work is in agreement with that of
other researchers (Stevenson \etal 1985; Pritchet \& Infante
1992; Jones, Shanks, \& Fong 1988; Bernstein \etal 1994; Brainerd, Smail,
\& Mould 1995 ), who found that the scaled two-point angular
correlation function amplitude is consistent with that found
locally, except for hints of mild evolution in luminosity and/or
clustering.

     Given the unexpected nature of our results (see below) we
have performed numerous checks to determine the validity of our
measurements.  It should, however, be emphasized that these data
are derived from multiple plates and that the results are
qualitatively identical for the three independent color
combinations for both red and blue galaxies in two separate
fields. The similarity of results decreases the probability that
artifacts, such as gradients across the plates in different
passbands, are responsible for the signal. As a further check
against artifacts caused by gross variations in color over the
plates, the correlation analyses were rerun with the median
$U-R_F$ color equilibrated over the plates by use of a 400"x400"
square smoothing window. Although this adjustment, which amounted
to several tenths of magnitude in some areas, weakened the
signals, all general trends were still clearly evident.
Additional tests were run on the data to check whether either
population was associated with bright objects such as galaxies or
stars in the fields. These proved negative. Cross correlations
between stars and galaxies in the fields also indicated no
obvious problems with the catalogs as did analysis using a subset
of stars identified through proper motion analysis.

\heading{3. Results}

\subheading{3.1 Measured Angular Correlation Amplitudes}

     The correlation amplitudes for SA57 and SA68, as a function
of $U- R_F$ color, are shown in Figure 1, together with plots
showing the percentage of data included in each subset. These
results show the angular correlation amplitude of the reddest and
bluest subsets of galaxies, respectively. The horizontal line is
the correlation amplitude of the whole sample.  In order to
present a single number, an amplitude, the correlation function
was integrated between $9$ and $144$ arcsecs to reduce noise and
for compatibility with the earlier work (Koo \& Szalay 1984).
Assuming a nominal redshift of $0.3$, $H_o$ = 50 km
sec$^{-1}$Mpc$^{-1}$, and $q_o \sim 0$ these limits span 70 kpc
to 1.1 Mpc. The amplitude of \wt, $A_w$, is reported using the
value of $-0.8$ for the power law slope at one degree in units of
$10^{- 4}$. For the average of the two fields we obtained
$A_w({\rm all})=18.0\pm1.8$.  Statistical (Poisson) error bars
are shown for all results, although these certainly underestimate
the true variance in the signal, as will be discussed below.

     Both fields show a marked enhancement in correlation
amplitude, with increases greater than a factor of ten for the
extreme subsets in both red and blue. As a check against the
possibility that only the most extreme galaxies contribute to the
rise in amplitude (because of artifacts in the data), we calculated
$w(\theta)$ in bands of color for both the red and blue galaxies.
The results are shown in Figure 2. It is evident that the
increase in angular correlation function exists over more than
one magnitude in color, which indicates that the enhancement is
genuine.

     Figure 3 shows \wt~ for `red' galaxies $U-R_F > 1.5$ and
`blue' galaxies $U-R_F<-0.25$ together with the entire data set
for both fields in five bins of equal logarithmic width between 18" and
289".  These results are compared against a fiducial
no-evolution model assuming a power law index of -0.8,
$q_o$=0.01, and $H_o$=50 km/sec.  Both the red and blue subsets
in each field show qualitative agreement with a power law slope
of -0.8, together with enhanced clustering over this entire
range.

     Other recent correlation function studies (Efstathiou \etal
1991; Brainerd \etal 1995; Neuschaefer \etal 1995) that used CCD
detectors have found quite weak angular correlation functions for
objects with an isophotal magnitude $24< B_J <26$. When the
Efstathiou \etal (1991) sample was divided into red and blue
subsets, the blue subset was found to have a marginally
($<2\sigma$) stronger correlation amplitude. One of their fields,
SA68, overlaps with one of ours. When we apply a similar color
split to our data as adopted by these authors, we obtain similarly
low amplitudes (as seen on Figure 1). Only with more extreme
color cuts do the enhancements in correlation amplitude become
evident.

     Our error bars correspond to the statistical noise in the LS
estimator (Landy \& Szalay 1993), i.e. we assume that the
distribution of the total number of galaxies in the field is
Poisson, thus the variance in the correlation is Poisson in the
bin counts. The LS estimator, conditional on the number of
galaxies observed, is optimal in this case, and still minimizes
the variance in all the other cases. If, however, there is
additional large scale clustering (as shown by the presence of a
cluster), the irreducible higher order moments of the galaxy
counts on scales of the whole field will result in additional
variance. An {\it a priori} determination of this variance, on the
other hand, requires knowledge of these higher order moments at
these magnitudes, colors, and angular scales. As these are not
available, we only plot the Poisson errors, but expect the full
error to be about a factor of two higher.
\smallskip
\centerline{\bf Table 1}
\medskip\hbox to \hsize{\hfill
  \vbox{\tabskip=0pt \offinterlineskip
    \def\tablerule{\noalign{\hrule}}
    \def\myspace{\omit&\multispan4&\cr}
    \halign {\strut#&
     &\hfil#\hfil &\hfil#\hfil&\hfil#\hfil\tabskip=0pt \cr
     \tablerule\myspace\myspace
     \tablerule\myspace
     &\multispan4 \hfil{
 Auto and Cross-Correlations by Bands of Color}\hfil\cr
     \myspace\tablerule\myspace
&\q{}    &  \q{Blue (20\%)}    &   \q{Mid (60\%)}  &   \q{Red
(20\%)}    \cr
     \tablerule\myspace\myspace\myspace
&\hfill Blue   &  68$\pm$ 20 &  9$\pm$ 7  &  -18$\pm$ 10 \cr
&\hfill Mid    &             & 14$\pm$ 1  &   20$\pm$  9 \cr
&\hfill Red    &             &            &  130$\pm$ 30 \cr
     \myspace\myspace\myspace\tablerule\cr}
}\hfill}

     Table 1 reports the auto- and cross-correlations of the top
20\% blue and red galaxies along with the mid 60 \%. The error
bounds in the table are derived from the actual scatter between the
two fields, and appear to be twice the statistical, Poisson
errors. These results show that the auto-correlations of the
intermediate colors are weaker than for the whole data set, and
the strong correlations in the two extreme colors boost the
overall clustering amplitude to its observed value $18.0\pm1.8$.
In addition, the cross correlation function between the red and
blue populations is marginally anti-correlated.

\subheading{3.2 Comparison with Models}

     We have compared our results to predictions of models
(Gronwall \& Koo 1995), designed to fit the observed counts,
colors, and redshift distributions, but that rely on having the
luminosity function of field galaxies be a free ``parameter'',
i.e., one that is not constrained by local observations.  The
models include some mild evolution in the stellar populations and
a small amount of intrinsic absorption/reddening due to dust. The
use of models was not meant to be part of an exhaustive analysis
of various evolutionary and cosmological scenarios, but rather to provide
vital baseline estimates of the redshift distributions needed to
interpret the angular correlations. These models are improved
estimates over those of prior analyses of angular correlation
functions (Efstathiou et al 1991, Koo \& Szalay 1984), since they
are also found to fit a wide range of other faint galaxy counts,
colors, and redshifts.

     To compare the actual observations, simulated galaxy
catalogs were generated that account for photometric errors and
incompleteness that mimic those expected to exist in our data.
We then applied identical magnitude ($20<B_J <23.5$) and color
cuts ($U-R_F > 1.5$ and $U-R_F<-0.25$) to the simulated catalogs
and derived the redshift distributions. These were used to
calculate the angular correlation amplitude, assuming a standard
no-evolution, -1.8 slope correlation function. The models
reproduce the overall correlation amplitude accurately (18.2),
and can be reconciled with the amplifications seen in the `red'
subsets, but fail to explain the angular clustering of the
extremely `blue' subset,
$$\eqalign{
     A_w({\rm blue})/A_w({\rm all}) =& 0.44 \cr
     A_w({\rm red})/A_w({\rm all})  =& 1.67 \cr
     A_w({\rm cross})/A_w({\rm all}) =& 1.02 \cr
}\nxt$$
The discrepancy is so large that no minor modification of the
models can give a satisfactory agreement. The redshift
distribution for the `blue' galaxies was more extended than for
the whole sample (because of evolution), resulting in a larger number
of projected pairs, thus in a lower amplitude for the predicted
angular correlation amplitude. Thus the  main question is
what constraints these observations generate on the redshift
distribution of the bluest galaxies.

\subheading{3.3 Comparison to Redshift Surveys}

     Deep redshift surveys are just reaching our magnitude limits
(Colless \etal 1991, 1993; Glazebrook \etal 1995;
Lilly \etal 1995).  The redshift survey most similar in limiting
magnitude to our data is the Glazebrook \etal (1995) survey to
$B<24$. This sample has 73 measured redshifts. Taking the 21
bluest galaxies of the $B<23.5$ subset, only 10 have redshifts;
thus the incompleteness is over 50\%, whereas the whole survey was
$\approx 70$\% complete. These 10 galaxies were rather evenly
distributed over the redshift range of 0.08 to 1.1; thus this blue
subset of the spectroscopic sample is too small and incomplete to
draw reliable conclusions about the detailed redshift distribution
to be drawn when such narrow color divisions are applied. The Lilly \etal
(1995) sample to $I<22$ has a larger number of redshifts (600),
but these are unpublished. We discuss their result in the Summary.

\heading{4. Discussion and Analysis}

\subheading{4.1 The Red Galaxies}

     In some respects, the large enhancement in the correlation
amplitude of the reddest galaxies may seem as surprising as that
for the blue subset. However, this result can be explained in
terms of known properties of red galaxies quite naturally.  An
enhancement in the correlation amplitude for the red galaxies may
be expected since these are primarily early type (E/S0) galaxies
(morphology-density relation). Further inspection of the red
subsets found in these fields reveals that field SA57 contains a
known cluster at $z=0.24$ and SA68 contains a supercluster at a
redshift of $z= 0.54$, both visible in maps of the fields (Figs.
10 and 11 in Koo 1986). From the models we have estimated that,
for galaxies redder than $U-R_F>1.5$ (with a median redshift of
0.34) the correlation amplitude is enhanced by a factor 1.67
from the modified redshift distribution alone.  Since that
population is hardly evolving, these model predictions should be
quite robust.  On the other hand, a 4 times stronger intrinsic
clustering has actually been measured for E/S0 galaxies (Nicotra
\etal 1993), resulting in a total correlation amplification of
6.4. Therefore, the measured overall enhancement of the red
galaxy signal can be well understood in these terms, and
illustrates the sensitivity of using angular correlations as a
function of color to study both galaxy evolution and large scale
structure.

\subheading{4.2 Blue Galaxies}

     Analyzing the enhancement in the correlation amplitude of
the bluest galaxies is more problematic than that of the red
subset since these have not been identified securely with a
well-studied local population. To reach quantitative conclusions
about their redshift distribution, we will exploit two basic
observational constraints on blue galaxies which generate
opposing trends: (1) the amplification of the very blue galaxy
angular correlations is greater than 6, and (2) the population
fraction is greater than 15\%. These constraints provide strong
limits on the possible redshift distributions.

     Several hypotheses have been proposed to explain the nature
and distribution of faint blue galaxies. One hypothesis is that
faint blue galaxies are the precursors of normal galaxies
but undergoing enhanced star formation at an earlier epoch, $z>1$.
Another is that these galaxies are a merging population, perhaps
at intermediate to high redshifts, since merging galaxies are
expected to be fairly bright and blue because of enhanced star
formation (Broadhurst, Ellis, \& Glazebrook 1992). Other
hypotheses include a dominant population of relatively nearby,
low surface brightness galaxies that are not represented in
existing estimates of the local luminosity function (McGaugh
1994) and the existence of a local bursting dwarf population with strong
luminosity
evolution that has presently faded from view (Broadhurst \etal
1988; Babul \& Rees 1992; Eales 1993; Lilly 1993). Below, we
discuss these possibilities to assess whether any of these
scenarios is consistent with our correlation function
measurements.

     An enhancement in autocorrelation amplitude for the blue
subsample can be explained either by these galaxies being
intrinsically more strongly correlated than commonly assumed, or
their redshift distribution being very different, or a
combination of both. Given that the intrinsic spatial correlation
of late-type galaxies has been found to be 4 times weaker
than that of early type galaxies (Nicotra \etal 1993), arguments that
postulate increased correlation amplitudes appear to be untenable. To
explain the observed enhancement in \wt\ solely by this effect
would require the opposite trend, namely, a 12 times stronger
intrinsic clustering for the very blue subset if they had the
redshift distribution as predicted by the model, or 6 times
stronger if they had the redshift distribution of the whole
sample. This leaves the possibility that the redshift
distribution of the bluest galaxies is quite different from the
predictions.  Since the angular correlation amplitude is
proportional to the second moment of $dn/dz$, the most obvious
implication of the enhanced clustering amplitude is that the
redshift distribution is narrower than the models suggest.

     In considering the possibility that the bluest galaxies are
all at high redshift, let us assume that our blue galaxies are
the distant counterparts of normal galaxies today that have
undergone enhanced star formation in the past. Having measured
the power-law slope of the correlation function of the blue
galaxies to be similar to that of galaxies today, and assuming
that these counterparts predominantly lie at high redshifts $z
\ge 1$, the most straightforward method to increase the
correlation amplitude would be to have a strongly peaked redshift
distribution. Such a peaked distribution, though in principle
possible is unlikely given the broad shape of the luminosity
function, unless special galaxy formation models are invoked,
such as by Babul and Ferguson (1995).  Furthermore, the redshift that
correspondes to the 15 percentile point in the high redshift tail
of a standard galaxy distribution is about $z=0.76$ at our
magnitude limit. Any subset of this tail will not have enough
galaxies to make up the observed 15\%.

     Another explanation is that the very blue galaxies are
predominantly a rapidly merging population at $z\ge 1$.  In such
a scenario, however, one might expect these galaxies to have a
correlation function with an excess of close pairs.  While
several studies support an increase of close pairs at the level
of $(1+z)^{3}$ (Zepf \& Koo 1989; Burkey \etal 1994; Carlberg,
Pritchet, \& Infante 1994; Yee \& Ellingson 1995), the fraction
of participating galaxies remains small, at the level of 10\% or
less.  More importantly, very close pairs do not appear to have
colors that differ significantly from typical field galaxies.
Since we also find that the very blue galaxy correlation function
has the same smooth power-law slope as that of the red galaxies,
without a break to higher amplitudes on small scales, we consider
the merger hypothesis an unattractive explanation. In any
case, it is not the existence of close pairs that accounts for
the enhancements in our signals.

     The other extreme possibility is that the blue galaxies are at
low redshift. Indeed, the enhanced correlations suggest this to
be the case --- in the integral that gives the angular correlation
amplitude, the $(dn/dz)^2$ term has a very large weight at low
redshifts, thus it is much easier to account for the excess
clustering by modifying the redshift distribution at the low $z$
end.  To quantify this result, without going into an overly
complicated modeling scheme, we will try to parameterize the
phenomenological redshift distribution of the blue galaxies by a
simple functional form.  This will show that, independent of the
particular details, the only possible solution is to have most of
these galaxies at redshifts $z<0.3$.

     For this simple parametrization we make the following
assumptions: (1) the number counts combined with redshift surveys
fix the redshift distribution of the whole sample, with
$20<B_J<23.5$, (2) the red galaxies in the sample do not evolve,
and their redshift distribution is adequately described by the
models, (3) a standard, non-evolving clustering amplitude is
used throughout.

     For our first experiment, we try to assign the blue galaxies
to as low a redshift as possible. We adopt the following
algorithm for selecting the redshift distribution of the blue
subset: we subtract the red galaxies from the overall redshift
distribution, creating a complementary sample.  This subset is
multiplied by a simple lowpass window, and the resulting
distribution is assigned to the bluest subset. The overall
population fraction and amplification of \wt\ of this subset are
calculated.  Satisfactory subsets are required to have a fraction
greater than 15\%, and an amplification greater than 6. Although
this recipe falls short of providing a full modeling of all the
color distributions, number counts, and correlations, it is
straightforward and incorporates some basic galaxy distribution
properties. The window to select the low-redshift galaxies is a
simple, sharp low-pass filter, i.e. all galaxies below that
redshift are assigned to have such blue colors. This is clearly
unrealistic, but this is how we can make the redshifts of the
blue galaxies the lowest possible. We find that the highest
redshift cutoff with such models is at $z=0.28$.

   This simple model is clearly in conflict with the redshift
surveys. Glazebrook \etal (1995) found that at least 4 of the 21
bluest galaxies were at redshifts beyond 0.6.  Adopting an
additional conservative constraint, that at least 25\% of this
bluest subset has to be at $z>0.8$, we can do another experiment.
We start with the bluest galaxies from the models. Their redshift
distribution is quite flat. We add an extra low redshift
component from the rest of the galaxies, with a sharp upper
cutoff in their redshifts. The highest cutoff satisfying these
constraints is at $z=0.2$, in which case 60\% of the galaxies are
below $z=0.3$.  In this scenario all of the remaining galaxies
had to be assigned to the low redshift tail. The other extreme
case in such a picture is to assign extra galaxies at very low
redshifts, at $z<0.05$. In this case we can get away with adding
a small fraction of all the blue galaxies; however, the small
fraction necessary for the amplification exceeds the number of
available galaxies at that redshift by a factor of 2.

\heading{5. Conclusion: Low-Redshift Faint Galaxies}

     We conclude from these exercises that it is not possible to
explain the observed excess correlations with even moderately
large modifications of the class of models we have considered,
that is, a substantial fraction of the blue galaxies must exist
below a redshift of $z=0.3$.  However, such analyses alone cannot
discriminate between a population of blue galaxies all at
redshifts up to 0.3 (e.g. low surface brightness galaxies),
versus a strongly evolving and thus more peaked population.
However, we can certainly rule out the possibility that the bulk
of these galaxies are of high luminosities with a wide spread over
high redshifts. From the results of the data and the models, we
conservatively predict that more than 50\% of the currently
unidentified population of the faint blue galaxies at our
magnitude limit will be found at redshifts $z<0.3$.

     With evidence that these very blue galaxies are mostly
nearby, their luminosities can be estimated by applying a
K-correction of 0. The galaxies in this survey typically have $22
< B_J < 23.5$, which translates to limits on absolute magnitudes
of $-18 > M_{B_J} > -19.5$ at redshifts of 0.3. For lower
redshifts, the derived luminosities will, of course, be fainter.
Typical (L$^*$) galaxies have $M_J \sim -21.0$ and thus the very
blue galaxies appear to be of low luminosities.  If their blue
color is due to enhanced star formation activity, so that their
luminosity has increased as well, their progenitors may well be
from the faint end of the luminosity function.

     Since a rapid luminosity evolution of these blue galaxies,
combined with our magnitude cutoff at $B_J=23.5$, can create a
reasonably peaked redshift distribution between 0.2 and 0.3, a low
redshift population would result if the faint end slope of the
luminosity function of very blue galaxies were steep, as
suggested by the results of Marzke \etal (1994) or Metcalfe \etal
(1991). Such a hypothesis is not only supported by our work, but
also by recent redshift surveys. Consider the expected color
tracks traversed by galaxies of different intrinsic colors as a
function of redshift in the Canada-France Redshift Survey (see
for example Fig 2 of Lilly \etal (1995) which shows $(V-I)_{AB}$ vs
redshift to $I<22$).  The exact distribution of these galaxies
involves a complex interplay between the depth of the survey,
luminosity functions, evolution, cosmology, dust; however,
the bluest galaxies are found to be at either very low or very
high redshifts. Based on this figure of almost 600 galaxies, we
find that the bluest galaxies indeed appear to be concentrated at
low redshifts. More specifically, reading off the uncrowded
bluest 18 points, 11 are below $z = 0.3$, just as we have
predicted with our correlation analysis. The strongly evolving
luminosity function of the bluest half in this data set also
suggests that these effects are even more pronounced for the
bluest 15\%. Therefore, we conclude that the faint end slope of
the local luminosity function for blue galaxies is steeper than
that adopted by our baseline model and that perhaps the luminous
component at high redshifts is less. Thus this redshift survey
supports our work and shows the effectiveness of our approach.
Of the 10 bluest galaxies with redshifts in the Glazebrook \etal
(1995) sample, 4 had redshifts below $z=0.3$, and 4 were above
0.6. The remaining 11 are a large enough percentage that their
result may still be consistent with ours. We predict that these
remaining objects will be found to be low-metallicity systems
with redshifts less than 0.3.

     Having identified the blue galaxies as being dominated by a
low redshift population, we can estimate the expected strength of
the cross-correlation between the red and blue subsets. From the
models -- with the assumption that the intrinsic clustering is
average -- we obtain expected amplifications between 0.7 and 1.3,
which yields an amplitude range of $18\pm15$. Contrary to this
expectation is the observed value $-18\pm10$ shown in Table 1, a
discrepancy of $2\sigma$. However, our null hypothesis of
average clustering is likely to be incorrect; unfortunately no
local information is available on the intrinsic, spatial
cross-correlations of different galaxy types.  This measured
amplitude is marginally consistent with a null signal. If the
numbers are taken literally, we may speculate that the weak
anti-correlations indicate that only a small fraction of the very
blue galaxies can co-exist with the red galaxies at similar,
moderate redshifts, and for those that do co-exist, they occupy
separate environments. Since red, early type galaxies populate
high density regions, the faint blue galaxies should be in
surrounding, segregated lower-density regions, which is not inconsistent
with common beliefs. The primary limitation in this analysis,
which prevented us from being able to make much stronger
statements about the redshift distribution of these faint
galaxies, is the poor knowledge of the local auto-and cross
correlation properties of the different galaxy types.

     We have shown that angular correlations as a function of
color can serve as a very sensitive statistical tool to study
galaxy evolution as well as to provide new insight into the
nature of galaxies of extreme colors. Since multicolor
photometric observations easily reach beyond the practical limits
of reliable redshift measurements, angular correlation functions,
combined with color cuts, will play an increasingly important
role in understanding faint galaxies in the future. As the local
galaxy population is better understood, the accuracy of this
method will increase substantially.

\bigskip\bigskip
\centerline{\bf Acknowledgements}
\bigskip

The authors thank the several referees including Karl Glazebrook
for useful suggestions, C. Gronwall for
providing the model catalogs for our simulations, S. Majewski for
providing various stellar catalogs for checking against systematic
effects, and N. Roche for useful discussions. David Koo and Alex
Szalay have been supported by an NSF/Hungary exchange grant, and
by  a grant from the US-Hungary Joint Fund.  Alex Szalay has been
supported  by NSF grant AST 90-20380 and David Koo by AST
88-58203. Stephen Landy has been supported both by the Johns
Hopkins University and an NSF grant at Carnegie Observatories.

\centerline{\bf References}

  \def\ps{\noindent\goodbreak
     \parshape 2 0truecm 15truecm 2truecm 13truecm}
  \def\ref#1;#2;#3;#4;#5.{\ps #1\ #2, #3,\ #4, #5}

  \def\ApJ{ApJ}
  \def\ApJS{ApJS}
  \def\MNRAS{MNRAS}

\def\aj{AJ}
\def\pq{\par\noindent\parshape 2 0truecm 14.5truecm 2truecm
12.5truecm}
\def\apj#1;#2;#3;#4;#5{#1 #2, #3, #4, #5}
\def\pp{\pq\apj}
\def\bk{\hfil\break}
\parskip=0pt
\bigskip

\pp Babul,A. \& Rees,M.J.;1992;\MNRAS;255;346
\pp Babul,A. \& Ferguson,H.;1995;\ApJ;;submitted
\pp Bernstein, G. M., Tyson, J. A., Brown, W. R., \& Jarvis, J.F.;1994;
	\ApJ;426;516
\pp Brainerd,G.T., Smail,I., \& Mould,J.R.\bk
	;1995;\MNRAS;submitted;.
\pp Broadhurst,T.J., Ellis, R.S. \& Shanks,T.;1988;\MNRAS;235;827
\pp Broadhurst,T.J., Ellis, R.S. \& Glazebrook,K.;1992;Nature;355;55
\pp Burkey,J.M., Keel,W.C., Windhorst,R.A., \& Franklin,B.E.;1994;
	\ApJ;150;L13
\pp Carlberg,R.G., Pritchet,C. J.,\& Infante, L.;1994;\ApJ;435;540
\pq Colless,M.M. 1994, in Wide Field Spectroscopy and the Distant
     Universe, ed. S. Maddox, World Scientific, 263.
\pp Colless,M.M., Ellis,R.S., Broadhurst,T.J.,Taylor, K. \bk
	\& Peterson,B.H.;1993;\MNRAS;261;19
\pp Colless,M.M., Ellis,R.S., Taylor,K., \& Shaw,G.;1991;\MNRAS;253;686
\pp Cowie,L.L.,Songaila,A. \& Hu,M.;1991;Nature;354;460
\pp Eales,S.;1993;\ApJ;404;51
\pp Efstathiou, G.P.E.;1995;\MNRAS;272;L25
\pp Efstathiou,G.P.E., Bernstein,G., Katz,N., Tyson,J.A. \&
     Guhathakurta,P.;1991;\ApJ;380;L47
\pp Giovanelli,R., Haynes,M.P. \& Chincarini,G.L.;1986;\ApJ;300;77
\pp Glazebrook,K., Ellis,R.S., Colless,M.M., Broadhurst, T.J.,\\
	Allington-Smith,J.R., Tanvir,N.R. \& Taylor,K.;
	1995;\MNRAS;273;157
\pp Griffiths,R.E., Casertano,K.U., Ratnatunga,L.W., Neuschaefer,L.W.,\bk
	Ellis,R.S., Gilmore,G.F., Glazebrook,K., Santiago,B., Huchra,\bk
	J.P., Windhorst,R.A., Pascarelle,S.M., Green,R.F.,\bk
	Illingworth,G.D., Koo,D.C. \& Tyson,A.J.;1994;\ApJ;435;L19
\pp Gronwall,C. \& Koo,D.C.; 1995;\ApJ;440;L1.
\pp Infante,L. \& Pritchet,C.J.;1992;\ApJS;83;237
\pp Infante,L. \& Pritchet,C.J.;1995;\ApJ;439;565
\pq Jones,L.R., Shanks,T., Fong,R. 1988,\bk in
    Large Scale Structures of the Universe,\bk
    IAU Symp. 130 ed. J. Audouze, M.C. Pelletan, A.S. Szalay \bk
  (Dordrecht:Kluwer), 528.
\pq Koo, D. C. 1996, in Examining the Big Bang and\bk
    Diffuse Background Radiations,IAU Symp. No. 168,\bk
    ed. M. Kafatos (Kluwer: Dordrecht), in press
\pp Koo, D.C. ;1986;\ApJ;311;651
\pp Koo, D.C., \& Kron, R. G. A.;1992;Rev. ARA\&A;30;613
\pp Koo, D.C., \& Szalay, A. S. ;1984;\ApJ;282;390
\pp Kron, R.G.;1980;\ApJS;43;305
\pp Landy, S.D., \& Szalay, A.S.;1993;\ApJ;412;64
\pp Lilly, S.J.;1993;\ApJ;411;501
\pq Lilly, S.J., LeFerre, O., Hammer, F., Crampton, D., \& Tresse,L.\bk
	1995, in Wide Field Spectroscopy of the Distant Universe,\bk
	(Singapore:World Scientific)
\pp Marzke,R.O., Geller,M.J., Huchra,J.P. \& Corwin,H.G.;1994;\aj;108;437
\pp Metcalfe,N., Shanks,T., Fong,R.\& Jones,L.R.;1991;\MNRAS;249;498
\pp McGaugh,S.;1994;Nature;367;538
\pp Neuschaefer,L.W., \& Windohirst,R.A.;1995;\ApJ;439;14
\pq Nicotra,M.A., Abadi,M.G.,\& Lambas,D.G. 1993, in Observational Cosmology\bk
	ed. G.Chincarini, A.Iovino, T.Maccacaro and D. Maccagni\bk
	(ASP Conference Series, Vol 51 1993), 152.
\pp Neuschaefer,L.W., Ratnatunga,K.U., Griffiths,R.E., Casertano,S. \\
	\& Im,M.;1995;\ApJ;;{in preparation}
\pq Peebles,P.J.E. 1980 in The Large Scale Structure of the Universe,\bk
    (Princeton Univ. Press, Princeton)
\pp Pritchet,C.J. \& Infante,L.;1992;\ApJ;399;L35
\pp Stevenson,P.R.F., Shanks,T., Fong,R. \& MacGillivray,H.T.\bk
     ;1985;\MNRAS;213;953
\pp Yee,H.S. \& Ellingson,E.;1995;\ApJ;;{in press}
\pp Zepf,S.E. \& Koo,D.C.;1989;\ApJ;337;34

\end